\documentclass[12pt]{amsart}

\usepackage{amsmath}
\usepackage{amsfonts}
\usepackage{amssymb}
\usepackage{latexsym}

 \advance\hoffset by -0.4 truecm

\newtheorem{lemma}{Lemma}
\newtheorem{corollary}{Corollary}

\newtheorem{theorem}{Theorem}

\newcommand{\s}{\vspace{0.2cm}}
\newcommand{\R}{\text{\rm Re }}
\newcommand{\I}{\text{\rm Im }}

\def\Vb{\text{\bf V}}

\newcommand{\Diff}{\text{\rm Diff }}
\newcommand{\Rot}{\text{\rm Rot }}
\newcommand{\Vect}{\text{\rm Vect }}

\begin{document}
\title[Liouville and logarithmic actions...]
{Liouville and logarithmic actions in Laplacian growth}
\author[Alexander Vasil'ev]{Alexander Vasil'ev}

\address{Matematisk institutt, Universitetet i Bergen, Johannes Brunsgate 12, N-5008 Bergen, Norway}
\email{alexander.vasiliev@uib.no} 
\begin{abstract}
We discuss and construct an action functional (logarithmic action)
for the simply connected Laplacian growth and obtain its variation. This variation admits 
various interpretations. In particular, we consider a general smooth subordination evolution and give connections with the Virasoro algebra and Neretin polynomials. 
\end{abstract}

\maketitle

\section{Introduction}

An important quantity in mechanics and the field theory is the {\it
action}, a local functional over maps $\phi$ between the space-time
$\Sigma$ and the target space which we assume to be the Euclidean
straight line $\mathbb R$. For example, one may consider the action
given by the Dirichlet integral
\[
\mathcal{S}[\phi]=\int\limits_{\Sigma}|\nabla \phi|^2 dv,
\]
where the metric structure of $\Sigma$ and the volume element $dv$
are to be taken into account as well as certain properties of
smoothness of $\phi$, that give sense to the right-hand side of the
above equality. In the classical setup of mechanics this action
represents the energy of the system as infinitesimally it is just
the scalar product of the field $\nabla \phi$ and its momentum
$\nabla \phi \,dv$ for the potential $\phi$. We note that in
general, the action is the integral over the classical Lagrangian.
If time does not enter explicitly into the Lagrangian, then the
system is closed, and a typical example of such a Lagrangian is the
kinetic energy minus the potential energy. One may assume different
smooth functions on $T\Sigma$ as Lagrangians, however physical or
geometrical background of the underlying space $\Sigma$ forces
certain restrictions in the choice. Nevertheless, different points
of view on the same object can lead to different functionals as
Lagrangians.

The classical field theory studies the extremum of the action
functional, and its critical value is called the {\it classical
action}. The critical point $\phi^*$ satisfies  Hamilton's principle
(or the {\it principle of the least action}), i.e., $\delta
\mathcal{S}[\phi^*]=0$, which is the Euler-Lagrange equation for the
variational problem defined by the action functional. For the action
given by the Dirichlet integral the classical action is achieved for
the harmonic $\phi^*$ and the principle of the least action leads to
the Laplacian equation $\Delta \phi=0$.

The {\it Liouville action} plays a key role in two-dimensional
gravity. It is based on a Lagrangian given on the base of the Riemannian
geometry of the underlying space. Liouville action describes a
highly non-trivial dynamics in quantum field theory and appears in
connection with Feynman's path integral that represents the
transition amplitude between two quantum states of a system
expressed as a sum over contributions from possible classical
histories of that system. Complex transition functions appear
naturally in the theory of evolution equations. The original
formulation of quantum Liouville theory through path integral has
been obtained by Polyakov \cite{Polyakov} in 1981 where the domain
of integration consisted of all smooth conformal metrics $ds^2$ on
an $n$-punctured Riemann sphere ($n\geq 4$ to guarantee the
hyperbolicity). A thorough mathematical treatment has been made
later by Takhtajan and Zograf (see \cite{T1}, \cite{Zograf},
\cite{ZT1}, \cite{ZT2}).

It is not very surprising that several ``quantum features" appear in
non-linear problems of hydrodynamics, in particular, in the {\it Laplacian Growth}
problem. In 1898 Hele-Shaw \cite{Hele} proposed his
famous cell that was a device for investigating a flow of viscous
fluid in a narrow gap between two parallel plates.

The  dimensionless model of a moving viscous incompressible fluid in
the Hele-Shaw cell is described by a potential flow with the
velocity field $\Vb=(V_1, V_2)$. The pressure $p$ is the potential
for the fluid velocity $$\Vb=-\frac{h^2}{12\mu}\nabla p,$$ where $h$
is the cell gap and $\mu$ is the viscosity of the fluid (see, e.g.
\cite{Ock, Saffman}).

Given an incompressible advancing fluid injected through a point
source, the Laplacian growth is formulated as a moving boundary
problem for the Laplacian equation for a function $p(z,t)$,
supported in a domain $\Omega(t)\subset \mathbb C$ as a function of
$z$,
\begin{equation}
\Delta p=-2\pi\delta_0(z),\quad z\in \Omega(t),\label{1}
\end{equation}
 where $t$ is the time
variable and $\delta_0$ is the Dirac distribution supported in $0$.
The dynamic boundary condition is given by putting
\begin{equation}
p\bigg|_{\partial \Omega(t)}=0, \label{2}
\end{equation}
and the kinematic condition for the motion of the boundary $\partial
\Omega(t)$ is given by the normal velocity  as
\begin{equation}
v_n=-\frac{\partial p}{\partial n},\quad x\in \partial
\Omega(t),\label{3}
\end{equation}
where $n$ is the outward unit vector to $\partial\Omega(t)$.

 Through the similarity in the governing
equations, Hele-Shaw flows can be used to study  models of saturated
flows in porous media governed by Darcy's law. Over the years
various particular cases of such a flow have been considered.
Different driving mechanisms were employed, such as surface tension
or external forces (suction, injection). We mention here a 600-paper
bibliography of free and moving boundary problems for Hele-Shaw and
Stokes flows since 1898 up to 1998 collected by Gillow and Howison
\cite{Gill}.

As it has been shown in \cite{Agam, Kostov, Marshakov, Wiegmann},
the Laplacian growth problem can be embedded into a larger hierarchy
of domain variations (Whitham-Toda hierarchy) for which all
Richardson's complex moments \cite{Richardson} are treated as
independent variables (generalized time variables), and form an
integrable system. Finally, the Laplacian growth has been modeled in
the moduli space of Riemann surfaces \cite{Krichever}.

In the classical simply connected case of the Laplacian growth
without gravity the kinetic energy is given by the Dirichlet
integral for the pressure as a potential. However, given the
evolution of the phase domain in time as a closed system with a
Riemannian metric as the geometrical background, one may construct
the action functional based on a different Lagrangian. The idea of
the construction of the Liouville action gives us a way to derive an
action functional for the Laplacian growth. The aim of this paper is
to construct the {\it logarithmic action} for the Laplacian growth
and to obtain its variation. Then we shall study a general smooth subordination
and interpret the variation of the logarithmic action through the infinitesimal 
version of the action of the Virasoro-Bott group on the space of analytic univalent
functions.

The author thanks Professor Leon Takhtajan for discussions on this
subject.

\section{Liouville and logarithmic actions}

Following the classical approach by Poicar\'e \cite{Poincare} we
consider a compact Riemann surface $S$ of genus $g\geq 2$ that
admits uniformization by a Fuchsian group $G$ acting in the unit
disk $U$, $S=U/G$. If $z$ is a local analytic coordinate defined on
an open set, then the Riemann metric $ds^2$ is represented as
$ds^2=\rho^2 |dz|^2$ for a density $\rho$. A {\it conformal metric}
corresponds to the invariancy under the change of the local
coordinate. If $z'$ is another local coordinate defined in an open
set and $ds^2=(\rho')^2 |dz'|^2$ in terms of the local coordinate
$z'$, then  we have $\rho(z)\equiv \rho'(z'(z))|dz'/dz|$ in the
intersection of these sets.

The Gaussian (sectional) curvature $\varkappa$ of this metric is
calculated by the formula $\varkappa=-\frac{1}{\rho^2}\Delta
\log\,\rho$. It follows from the uniformization theorem that there
exists a unique conformal metric of Gaussian curvature
$\varkappa=-1$ which is called the Poicar\'e (hyperbolic) metric. In
terms of the parameter $z$ this is equivalent to the {\it Liouville
equation}
\begin{equation}
\varphi_{z\bar{z}}=\frac{1}{2}e^{\varphi},\label{Lequation}
\end{equation}
where $\varphi=\log\,\rho^2$. In the simplest case of the unit disk
the Poicar\'e metric is given as
\[
ds^2=\frac{4|d\zeta|^2}{(1-|\zeta|^2)^2}.
\]
 Considering the universal covering
of $S$ by the unit disk and the automorphic (with respect to $G$)
projection $h: U\to S$ we deduce that
\[
ds^2=\frac{4|{h^{-1}}'|^2|d\zeta|^2}{(1-|h^{-1}|^2)^2},
\]
where $\zeta$ is taken from the fundamental polygon $U/G$. This
metric is complete and the area of $S$ in this metric is $4\pi(g-1)$
(by the Gauss-Bonnet theorem).

It seems that equation (\ref{Lequation}) is the Euler-Lagrange
equation for the variational problem defined by the
 functional
\begin{equation}
S[\phi]=\int\limits_{U/G}(|\phi_z|^2+e^{\phi})d\sigma_z,
\label{Laction}
\end{equation}
where $d\sigma_z=\big|\frac{dz\wedge \, d\bar{z}}{2} \big|$ which
may be chosen as the {\it Liouville action}. However, the deal is
much more difficult, because the integrant (the first term) does not
keep the local 2-form $|\phi_z|^2 dz\wedge \, d\bar{z}$ invariant
under the change of the local parameter from chart to chart on the
Riemann surface. Thus, the functional (\ref{Laction}) is
well-defined only for simply connected domains and the correct
Liouville action requires an additional term which has been given by
Takhtajan and Teo in \cite{T2}.

When the underlying Riemann surface have singularities (e.g.,
punctures, branch points) the action functional is not well defined
globally on the surface either. In this case, there are two options
to treat the problem. One of them is to change the metric, the other
is to make certain regularization. Actually, Poincar\'e
\cite{Poincare} worked in the presence of parabolic singularities.
In this case some additional terms appear in (\ref{Laction}), see
\cite{ZT2}. But certainly one may look for equilibrium between these
two options.

\s

Let us turn now to the Laplacian growth. As it was said in
Introduction, the problem (\ref{1}--\ref{3}) defines the Laplacian
growth. More thoroughly we give a {\it strong formulation} of this
problem. Let $\Omega(t)\subset\mathbb C$, $0\in \Omega(t)$, be a
one-parameter family of bounded domains. We call the family
$\{\Omega(t)\}$ smooth if $\partial \Omega(t)$ are smooth
($C^{\infty}$) interfaces for each $t$, and the normal velocity
$v_n$ continuously depends on $t$ at any point of $\partial
\Omega(t)$. Each $\Omega(t)$ is supposed to be simply connected for
any $t\in [0,T)$ fixed. A smooth  family $\Omega(t)$, $0\leq t< T$,
as above, is said to be a strong solution for the Laplacian growth
if there exists a potential $p(z,t)$, $z\in \Omega(t)$, such that
all conditions (\ref{1}--\ref{3}) are satisfied. The family of
$\Omega(t)$ forms a strong subordination chain of bounded domains:
$\overline{\Omega(s)}\subset \Omega(t)$ for $0\leq s<t<T$, $0\in
\Omega(0)$.

It is known that if the initial domain $\Omega(0)$ has an analytic
smooth boundary, then the strong solution exists locally in time
until the boundary $\partial\Omega(t)$ develops a cusp in a blow-up
time or $\Omega(t)$ changes its topology (see, e.g. \cite{How,
Reissig, Vas, VinKuf}).

By the Riemann theorem we construct a conformal time-dependent map
$z=f(\zeta,t)$ from the unit disk $U$ onto the phase domain
$\Omega(t)$, $f(0,t)=0$, $f'(0,t)>0$. The function
$f(\zeta,0)=f_0(\zeta)$ parameterizes the initial
 boundary $\partial\Omega(0)=\{f_0(e^{i\theta}),\theta\in [0,2\pi)\}$
and the moving boundary is parameterized by
$\partial\Omega(t)=\{f(e^{i\theta},t),\theta\in [0,2\pi)\}$. We use
the notations $\dot f=\partial f/\partial t$, $f'=\partial
f/\partial \zeta$.

Let us consider the complex potential $W(z,t)$, $\R W=p$. For each
fixed $t$ it is a multivalued analytic function defined in
$\Omega(t)$ whose real part solves the Dirichlet problem
(\ref{1}--\ref{2}).
 Making use of the Cauchy-Riemann conditions
we deduce that
\begin{equation*}
\frac{\partial\,W}{\partial\,z}=\frac{\partial p}{\partial
x}-i\frac{\partial p}{\partial y},\quad z=x+iy.
\end{equation*}
Since Green's function solves (\ref{1}--\ref{2}), we have the
representation
\begin{equation}
W(z,t)=-\log z+w_0(z,t),\label{13}
\end{equation}
where $w_0(z,t)$ is an analytic regular function in $\Omega(t)$.
Because of the conformal invariance of Green's function we have the
superposition
\[(W\circ f)(\zeta,t)=-\log\,\zeta,\]
and the conformally invariant {\it complex velocity} is just
$W'(z,t)=-\frac{{f^{-1}}'}{f^{-1}}(z,t)$, where $\zeta=f^{-1}(z,t)$
is the inverse to our parametric function $f$ and prime means the
complex derivative. Rewriting this relation we get
\begin{equation}
(W'(z,t)\,dz)^2=\frac{d\zeta^2}{\zeta^2}.\label{qudratic}
\end{equation}
The velocity field $(-\nabla p)$ is the conjugation of $(-W')$. In
other words the velocity field is directed along the trajectories of
the quadratic differential in the left-hand side of (\ref{qudratic})
for each fixed moment $t$. The equality (\ref{qudratic}) implies
that the boundary $\partial\Omega(t)$ is the orthogonal trajectory
of the differential $(W'(z,t)\,dz)^2$ with a double pole at the
origin. The dependence on $t$ yields that the trajectory structure
of this differential changes in time, and in general, the stream
lines are not inherited in time. These lines are geodesic in the
conformal metric $|W'(z,t)||dz|$ generated by this differential.

Observe that being thought of as a Riemann surface, the phase domain
$\Omega(t)\setminus\{0\}$ is hyperbolic and it admits  the
Poincar\'e metric with constant negative curvature
\[
ds^2=\frac{|{f^{-1}}'|^2}{|f^{-1}|^2\log^2|f^{-1}|}|dz|^2.
\]
 The asymptotics about
the origin and close to the hyperbolic boundary implies that the
standard expression (\ref{Laction}) for the Liouville action can not
be used any more. Moreover, the action integral for the hyperbolic
metric in the punctured unit disk has the following asymptotics
\begin{eqnarray*}
\int\limits_{U_{\varepsilon}}(|\phi_z|^2+e^{\phi})d\sigma_z &\sim &
-2\pi\log\varepsilon_1-4\pi\log|\log\varepsilon_1|\nonumber\\&+&
4\pi\log|\log(1-\varepsilon_2)|-\frac{4\pi}{\log(1-\varepsilon_2)},
\end{eqnarray*}
where $U_{\varepsilon}=\{z:\,\varepsilon_1<|z|<1-\varepsilon_2\}$.
Therefore, the corresponding integral for $\Omega(t)\setminus\{0\}$
will have a similar asymptotics plus terms containing $f'(0,t)$ and
the boundary distortion (in our case the boundary derivatives) at
the unit circle that makes it difficult to operate with.

Let us use the flat {\it logarithmic metric} instead generated by
(\ref{qudratic}) which seems to be more natural for the Laplacian
growth
\[
ds^2=\frac{|{f^{-1}}'|^2}{|f^{-1}|^2}|dz|^2=|W'|^2|dz|^2.
\]
 The hyperbolic boundary
is not singular for this metric whereas the origin is. But it is a
parabolic singularity which can be easily regularized.

 The density of this
metric satisfies the usual Laplacian equation $\varphi_{z\bar{z}}=0$
in $\Omega(t)\setminus\{0\}$, where
$\varphi(z)=\log\frac{|{f^{-1}}'|^2}{|f^{-1}|^2}$. Obviously, the
Laplacian equation is the Euler-Lagrange equation for the
variational problem defined by the Dirichlet integral
\[
\int\limits_{D}|\phi_z|^2 d\sigma_z,
\]
locally for any measurable set $D\subset \Omega(t)\setminus\{0\}$.
However, this functional cannot be defined globally in
$\Omega(t)\setminus\{0\}$ because of the parabolic singularity at
the origin. To overcome this obstacle we define the classical action
in the following way. Let
$\Omega_{\varepsilon}(t)=\Omega(t)\setminus
\{z:\,|z|\leq\varepsilon\}$ for a sufficiently small $\varepsilon$,
$U_{\varepsilon}=\{\zeta:\,\varepsilon<|\zeta|<1\}$. The function
$\varphi$ possesses the asymptotics
\[
\varphi\sim \log \frac{1}{|z|^2},\quad  |\varphi_z|\sim
\frac{1}{|z|^2} \quad \mbox{as $z\to 0$}.
\]
Therefore, the finite limit
\begin{equation}
\lim\limits_{\varepsilon\to
0}\left\{\int\limits_{\Omega_{\varepsilon}(t)}|\varphi_z|^2
d\sigma_z+2\pi\log\varepsilon\right\}=:\mathcal{S}[\varphi]\label{action1}
\end{equation}
exists and we call it the {\it logarithmic action} for the Laplacian
growth.

\begin{lemma}
The Euler-Lagrange equation for the variational problem for the
logarithmic action $\mathcal{S}[\phi]$ is the Laplacian equation
$\Delta\phi=-4\pi\delta_0(z)$, $z\in \Omega(t)$, where $\delta_0(z)$
is the Dirac distribution supported at the origin, where $\phi$ is
taken from the class of twice differentiable functions in
$\Omega(t)\setminus \{0\}$ with the asymptotics $\phi\sim -\log
|z|^2$ as $z\to 0$.
\end{lemma}
\begin{proof}
Let us consider first the integral
\[
\mathcal{S}_{\varepsilon}[\phi]=\int\limits_{\Omega_{\varepsilon}(t)}|\varphi_z|^2
d\sigma_z=\int\limits_{\mathbb
C}\chi_{\Omega_{\varepsilon}(t)}|\varphi_z|^2 d\sigma_z,
\]
where $\chi_{\Omega_{\varepsilon}(t)}$ is the characteristic
function of $\Omega_{\varepsilon}(t)$. Then, due to Green's theorem,
\begin{eqnarray}
\lim\limits_{h\to
0}\frac{\mathcal{S}_{\varepsilon}[\phi+hu]-\mathcal{S}_{\varepsilon}[\phi]}{h}
& =& 2\int\limits_{\mathbb C}\chi_{\Omega_{\varepsilon}(t)}\R
\phi_z\overline{u_z}\,d\sigma_z\nonumber\\
&=& -\frac{1}{2}\int\limits_{\Omega_{\varepsilon}(t)}u\Delta\phi\,
d\sigma_{z}+ \frac{1}{2}\int\limits_{\partial
\Omega_{\varepsilon}(t)}u\frac{\partial \phi}{\partial
n}\,ds,\label{limit}
\end{eqnarray}
in distributional sense for every $C^{\infty}(\mathbb C)$ test
function $u$ supported in $\Omega(t)$. On the other hand, we have
$\partial\phi/\partial n\sim -2/\varepsilon$ as $\varepsilon\to 0$
and $u=0$ on $\partial \Omega(t)$. Therefore, the expression
(\ref{limit}) tends to
\[
-\frac{1}{2}\int\limits_{\Omega(t)}u\Delta\phi d\sigma_{z}-2\pi
u(0),
\]
as $\varepsilon\to 0$, and the latter must vanish, that is
equivalent to the Laplacian equation mentioned in the statement of the lemma.
Obviously, the logarithmic term in the definition of
$\mathcal{S}[\phi]$ does not contribute into the variation.
\end{proof}

Straightforward calculation gives
\[
\varphi_z=\frac{-1}{f'}\left(\frac{f''}{f'}+\frac{1}{\zeta}\right)\circ
f^{-1}(z,t.)
\]
Hence, the action $\mathcal{S}$ can be expressed in terms of the
parametric function as
\begin{equation}
\mathcal{S}[\varphi]\equiv
\mathcal{S}[f]=\lim\limits_{\varepsilon\to
0}\left\{\int\limits_{U_{\varepsilon}}\bigg|\frac{f''}{f'}+\frac{1}{\zeta}\bigg|^2d\sigma_\zeta+2\pi\log\varepsilon\right\}
+2\pi\log|f'(0,t)|, \label{action2}
\end{equation}
or adding the logarithmic term into the integral we obtain
\begin{equation}
\mathcal{S}[f]=\int\limits_{U}\left(\bigg|\frac{f''}{f'}+\frac{1}{\zeta}\bigg|^2-\frac{1}{|\zeta|^2}\right)d\sigma_\zeta
+2\pi\log|f'(0,t)|. \label{action3}
\end{equation}
The functional (\ref{action3}) resembles the {\it universal
Liouville action} defined by Takhtajan and Teo in \cite{T3,T4} for
quasicircles which is based on conformal maps from the unit disk and from
its exterior.

Observe that the classical kinetic energy $\mathcal{E}$ for the
harmonic potential $p$ is calculated by the Dirichlet integral
\[
\int\limits_{D}|p_z|^2 d\sigma_z=\int\limits_{D}|W'|^2 d\sigma_z,
\]
locally for any measurable set $D\subset \Omega(t)\setminus\{0\}$.
However, this integral again cannot be defined globally in
$\Omega(t)\setminus\{0\}$. Treating $\mathcal{E}$ in the same way as
$\mathcal{S}$ we come to the following finite limit
\begin{equation*}
\mathcal{E}:=\lim\limits_{\varepsilon\to
0}\left\{\int\limits_{\Omega_{\varepsilon}(t)}|W'|^2
d\sigma_z+2\pi\log\varepsilon\right\},
\end{equation*}
or in terms of the parametric function $f$
\[
\mathcal{E}=\mathcal{E}[f]=2\pi\log|f'(0,t)|.
\]
The latter expression allows us to think of $\mathcal{E}$ as a
capacity which exactly corresponds to the physical sense of
$\mathcal{E}$ as minimal energy.

\section{Variation of the logarithmic action}

The Laplacian growth problem (\ref{1}--\ref{3}) being rewritten for
the parametric time dependent function $f: U\to\Omega(t)$ admits the
form of the so-called {\it Polubarinova-Galina equation}, which is
in principle, a reformulation of the kinematic condition (see, e.g.,
\cite{How, Vas}). Namely, the function $f(\zeta,t)$ satisfies the
non-linear first-order partial differential equation
\begin{equation}
\R \left(\dot{f}\,\,\overline{\zeta f'}\right)=1,\quad
|\zeta|=1,\label{PG}
\end{equation}
with the initial condition $f(\zeta, 0)=f_0(\zeta)$. We denote by
$S_f$ the Schwarzian derivative
\[
S_f=\frac{f'''}{f'}-\frac{3}{2}\left(\frac{f''}{f'}\right)^2,
\]
and by
\[
\varkappa(\theta,t)=\frac{\R
\left(1+\frac{e^{i\theta}f''}{f'}\right)}{|f'(e^{i\theta},t)|}
\]
the curvature of the boundary $\partial\Omega(t)$ at the point
$f(e^{i\theta},t)$.

\begin{theorem} Let $z=f(\zeta,t)$ be the parametric function for
the Laplacian growth, $\mathcal{E}[f]$ be the kinetic energy, and
$\mathcal{S}[f]$ be the logarithmic action. Then
\[
\frac{d}{dt}\mathcal{E}[f]=\int\limits_{0}^{2\pi}\frac{1}{|f'(e^{i\theta},t)|^2}d\theta,
\]
and
\[
\frac{d}{dt}(\mathcal{S}[f]+\mathcal{E}[f])=2\int\limits_{0}^{2\pi}\varkappa^2(\theta,t)\,d\theta+
\int\limits_{0}^{2\pi}\frac{2}{|f'(e^{i\theta},t)|^2}\R(e^{2i\theta}S_f)d\theta.
\]
\end{theorem}

\begin{proof}
Making use of the Cauchy-Schwarz representation we extend this
equation into the unit disk
\begin{equation}
\dot{f}=\zeta f'p(\zeta,t),\label{PG2}
\end{equation}
where
\begin{equation}
p(\zeta,t)=\frac{1}{2\pi}\int\limits_{0}^{2\pi}\frac{1}{|f'(e^{i\theta},t)|^2}\frac{e^{i\theta}+\zeta}{e^{i\theta}-\zeta}\,d\theta.\label{carat}
\end{equation} Immediately, we obtain that
\[
\frac{d}{dt}\mathcal{E}[f]=\frac{d}{dt}2\pi\log|f'(0,t)|=\int\limits_{0}^{2\pi}\frac{1}{|f'(e^{i\theta},t)|^2}d\theta,
\]
and hence,
\begin{equation}
\frac{d}{dt}\mathcal{S}[f]=2\R\int\limits_{U}\left(\overline{\frac{f''}{f'}}+\frac{1}{\overline{\zeta}}\right)
\left((1+\zeta\frac{f''}{f'})p(\zeta,t)+\zeta
p'(\zeta,t)\right)'d\sigma_\zeta+\int\limits_{0}^{2\pi}\frac{1}{|f'(e^{i\theta},t)|^2}d\theta.\label{deriv}
\end{equation}
The integral in the first term of the right-hand side of
(\ref{deriv}) we rewrite by Green's theorem as
\[
I=\frac{-1}{2i}\int\limits_{S^1}\left(\overline{\frac{f''}{f'}}+\frac{1}{\overline{\zeta}}\right)\left((1+\zeta\frac{f''}{f'})p(\zeta,t)+\zeta
p'(\zeta,t)\right)\,d\bar{\zeta}-\frac{1}{2}\int\limits_{0}^{2\pi}\frac{1}{|f'(e^{i\theta},t)|^2}d\theta,
\quad S^1=\partial U,
\]
taking into account a singularity at the origin. Applying the
Cauchy-Schwarz formula to the first term in $I$ containing $p$ we
arrive at
\begin{eqnarray*}
2\,\R I&=&
\int\limits_{0}^{2\pi}\bigg|1+e^{i\alpha}\frac{f''(e^{i\alpha},t)}{f'(e^{i\alpha},t)}\bigg|^2\frac{d\alpha}{|f'(e^{i\alpha},t)|^2}\\
&+&
\R\int\limits_{0}^{2\pi}\left(\overline{1+e^{i\alpha}\frac{f''(e^{i\alpha},t)}{f'(e^{i\alpha},t)}}\right)e^{i\alpha}p'(e^{i\alpha},t)d\alpha
-\int\limits_{0}^{2\pi}\frac{1}{|f'(e^{i\theta},t)|^2}d\theta,
\end{eqnarray*}
or
\begin{eqnarray*}
\frac{d}{dt}\mathcal{S}[f]=2\,\R I+
\int\limits_{0}^{2\pi}\frac{1}{|f'(e^{i\theta},t)|^2}d\theta&=&
\int\limits_{0}^{2\pi}\bigg|1+e^{i\alpha}\frac{f''}{f'}\bigg|^2\frac{d\alpha}{|f'(e^{i\alpha},t)|^2}\\&+&
\int\limits_{0}^{2\pi}\R\left(1+e^{i\alpha}\frac{f''}{f'}\right)\R
e^{i\alpha}p'(e^{i\alpha},t)\,d\alpha\\&+&\int\limits_{0}^{2\pi}\I\left(1+e^{i\alpha}\frac{f''}{f'}\right)\I
e^{i\alpha}p'(e^{i\alpha},t)\,d\alpha.
\end{eqnarray*}
These equalities are thought of as limiting values making use of the
analyticity of $f$ on the boundary and $f'(\zeta,t)\neq 0$ for all
$\zeta$ in the closure of the unit disk. Let us denote in the latter
expression by $J_2$ the last integral and by $J_1$ the intermediate
one. We have,
\[
J_2=\int\limits_{0}^{2\pi}\I\left(1+e^{i\alpha}\frac{f''}{f'}\right)\I\frac{1}{2\pi}\int\limits_{0}^{2\pi}
\frac{1}{|f'(e^{i\theta},t)|^2}\frac{2e^{i\theta}e^{i\alpha}}{(e^{i\theta}-e^{i\alpha})^2}\,d\theta\,
d\alpha.
\]
Obviously,
\[
\frac{\partial}{\partial\theta}\left(\frac{e^{i\theta}+\zeta}{e^{i\theta}-\zeta}\right)=\frac{-2\,\zeta
i e^{i\theta}}{(e^{i\theta}-\zeta)^2},\quad \mbox{and}\quad
\frac{\partial}{\partial\theta}\left(\frac{1}{|f'(e^{i\theta},t)|^2}
\right)=\frac{2}{|f'(e^{i\theta},t)|^2}\I\frac{e^{i\theta}f''}{f'}.
\]
Integrating by parts  and applying the Cauchy-Schwarz formula again
we obtain
\[
J_2=-2\int\limits_{0}^{2\pi}\left(\I\left(1+e^{i\theta}\frac{f''}{f'}\right)\right)^2\frac{d\theta}{|f'(e^{i\theta},t)|^2}.
\]
Now we turn to $J_1$
\[
J_1=\int\limits_{0}^{2\pi}\R\left(1+e^{i\alpha}\frac{f''}{f'}\right)\R\frac{1}{2\pi}\int\limits_{0}^{2\pi}
\frac{1}{|f'(e^{i\theta},t)|^2}\frac{2e^{i\theta}e^{i\alpha}}{(e^{i\theta}-e^{i\alpha})^2}\,d\theta\,
d\alpha.
\]
Here we change the order of integration and get
\[
J_1=\int\limits_{0}^{2\pi}\frac{1}{|f'(e^{i\theta},t)|^2}\R\frac{1}{2\pi}\int\limits_{0}^{2\pi}
\R\left(1+e^{i\alpha}\frac{f''}{f'}\right)\frac{2e^{i\theta}e^{i\alpha}}{(e^{i\theta}-e^{i\alpha})^2}\,d\alpha\,
d\theta.
\]
Integrating by parts we obtain
\[
J_1=\int\limits_{0}^{2\pi}\frac{1}{|f'(e^{i\theta},t)|^2}\R\frac{-i}{2\pi}\int\limits_{0}^{2\pi}\frac{\partial}{\partial\alpha}
\R\left(1+e^{i\alpha}\frac{f''}{f'}\right)
\frac{e^{i\alpha}+e^{i\theta}}{e^{i\alpha}-e^{i\theta}}d\alpha\,d\theta.
\]
The Cauchy formula gives
\[
J_1=\int\limits_{0}^{2\pi}\frac{1}{|f'(e^{i\theta},t)|^2}\R\left(\frac{e^{i\theta}f''}{f'}+\frac{e^{2i\theta}f'''}{f'}-
\left(\frac{e^{i\theta}f''}{f'}\right)^2\right)d\theta,
\]
or
\[
J_1=\int\limits_{0}^{2\pi}\frac{1}{|f'(e^{i\theta},t)|^2}\R\left(\frac{1}{2}\left(1+\frac{e^{i\theta}f''}{f'}\right)^2+e^{2i\theta}S_f-\frac{1}{2}
\right)d\theta,
\]
where $S_f$ is the Schwarzian derivative.

Summing up all these integrals we come to the conclusion that
\[
\frac{d}{dt}\mathcal{S}[f]=\int\limits_{0}^{2\pi}\frac{1}{|f'(e^{i\theta},t)|^2}\R\left(\frac{3}{2}\left(1+\frac{e^{i\theta}f''}{f'}\right)^2+e^{2i\theta}S_f+
\frac{1}{2}
\right)d\theta-\int\limits_{0}^{2\pi}\frac{1}{|f'(e^{i\theta},t)|^2}d\theta.
\]

We observe that
\[
\varkappa(\theta,t)=\frac{\R
\left(1+\frac{e^{i\theta}f''}{f'}\right)}{|f'(e^{i\theta},t)|}
\]
is the curvature of the boundary $\partial\Omega(t)$ at the point
$f(e^{i\theta},t)$. Integration by parts implies
\[
\int\limits_{0}^{2\pi}\frac{1}{|f'(e^{i\theta},t)|^2}\left(\I\left(1+\frac{e^{i\theta}f''}{f'}\right)\right)^2d\theta
\]
\[
=
\frac{-1}{2}\int\limits_{0}^{2\pi}\frac{1}{|f'(e^{i\theta},t)|^2}\R\left(\frac{1}{2}\left(1+\frac{e^{i\theta}f''}{f'}\right)^2+e^{2i\theta}S_f-\frac{1}{2}
\right)d\theta.
\]
So
\begin{eqnarray*}
\frac{d}{dt}\mathcal{S}[f]&=&\frac{3}{2}\int\limits_{0}^{2\pi}\varkappa^2(\theta,t)\,d\theta-
\frac{3}{2}\int\limits_{0}^{2\pi}\frac{1}{|f'(e^{i\theta},t)|^2}\left(\I\left(1+\frac{e^{i\theta}f''}{f'}\right)\right)^2d\theta
\\& &+
\int\limits_{0}^{2\pi}\frac{1}{|f'(e^{i\theta},t)|^2}\R\left(e^{2i\theta}S_f+
\frac{1}{2} \right)d\theta-\int\limits_{0}^{2\pi}\frac{1}{|f'(e^{i\theta},t)|^2}d\theta\\
&=&(\frac{3}{2}+\frac{3}{8})\int\limits_{0}^{2\pi}\varkappa^2(\theta,t)\,d\theta-\frac{3}{8}\int\limits_{0}^{2\pi}\frac{1}{|f'(e^{i\theta},t)|^2}
\left(\I\left(1+\frac{e^{i\theta}f''}{f'}\right)\right)^2d\theta\\
& &+
\int\limits_{0}^{2\pi}\frac{1}{|f'(e^{i\theta},t)|^2}\R\left((1+\frac{3}{4})e^{2i\theta}S_f+
(\frac{1}{2}-\frac{3}{8})\right)d\theta-\int\limits_{0}^{2\pi}\frac{1}{|f'(e^{i\theta},t)|^2}d\theta.
\end{eqnarray*}
Repeating this step we get
\begin{eqnarray*}
\frac{d}{dt}\mathcal{S}[f]
&=&\frac{3}{2}(1+\frac{1}{4}+\dots+\frac{1}{4^n})\int\limits_{0}^{2\pi}\varkappa^2(\theta,t)\,d\theta-
\frac{3}{2}\frac{1}{4^n}\int\limits_{0}^{2\pi}\frac{1}{|f'(e^{i\theta},t)|^2}
\left(\I\left(1+\frac{e^{i\theta}f''}{f'}\right)\right)^2d\theta\\
& &+
\int\limits_{0}^{2\pi}\frac{1}{|f'(e^{i\theta},t)|^2}\R\left((1+\frac{3}{4}+\dots+\frac{3}{4^n})e^{2i\theta}S_f+
(\frac{1}{2}-\frac{3}{8}-\dots-\frac{3}{2}\frac{1}{4^n})\right)d\theta\\
& &-\int\limits_{0}^{2\pi}\frac{1}{|f'(e^{i\theta},t)|^2}d\theta,
\end{eqnarray*}
at $n$-th iteration. Taking limit as $n\to\infty$, we finally obtain
\[
\frac{d}{dt}\mathcal{S}[f]=2\int\limits_{0}^{2\pi}\varkappa^2(\theta,t)\,d\theta+
\int\limits_{0}^{2\pi}\frac{2}{|f'(e^{i\theta},t)|^2}\R(e^{2i\theta}S_f)d\theta-\int\limits_{0}^{2\pi}\frac{1}{|f'(e^{i\theta},t)|^2}d\theta,
\]
as claimed in the statement of the theorem.
\end{proof}

In the simplest case of the circular evolution
$f(\zeta,t)=\sqrt{2t}\zeta$ we have
$\frac{d}{dt}\mathcal{S}[f]=\frac{d}{dt}\mathcal{E}[f]=\frac{\pi}{t}$.

\section{Parametric manifold $\Diff S^1/\Rot S^1$}

As it has been mentioned in Section 2, starting with a smooth
($C^{\infty}$) initial boundary $\partial\Omega(0)$ the classical
evolution of Laplacian growth is given by domains $\Omega(t)$ with
 smooth boundaries $\partial\Omega(t)$ as long as the classical
solution exists. Our aim now is to give an embedding of this
evolution into the parametric Kirillov's space $\Diff S^1/\Rot S^1$.

We denote the Lie group of $C^{\infty}$ sense preserving
diffeomorphisms of the unit circle $S^1=\partial U$  by $\Diff S^1$.
Each element of $\Diff S^1$ is represented as $z=e^{i\phi(\theta)}$
with a monotone increasing, $C^{\infty}$ real-valued function
$\phi(\theta)$, such that $\phi(\theta+2\pi)=\phi(\theta)+2\pi$.
 The Lie algebra for $\Diff S^1$ is
identified with the Lie algebra $\Vect S^1$ of smooth ($C^{\infty}$)
tangent vector fields to $S^1$ with the Poisson - Lie bracket given
by
$$[\phi_1,\phi_2]={\phi}_1{\phi}'_2-{\phi}_2{\phi}'_1. $$
Fixing  the trigonometric basis in $\Vect S^1$ the commutator
relations take the form
\begin{eqnarray*}
\left[\cos\,n\theta, \cos\,m\theta\right] & = &
\frac{n-m}{2}\sin\,(n+m)\theta+ \frac{n+m}{2}\sin\,(n-m)\theta,\\
\left[\sin\,n\theta, \sin\,m\theta\right]& = &
\frac{m-n}{2}\sin\,(n+m)\theta+ \frac{n+m}{2}\sin\,(n-m)\theta,\\
\left[\sin\,n\theta, \cos\,m\theta\right] & = &
\frac{m-n}{2}\cos\,(n+m)\theta- \frac{n+m}{2}\cos\,(n-m)\theta.
\end{eqnarray*}
There is no general theory of infinite dimensional Lie groups,
example of which is under consideration.  The interest to this
particular case comes first of all from the string  theory where the
Virasoro (vertex) algebra appears as the central extension of $\Vect
S^1$ (see Section 7). The central extension of $\Diff S^1$ is called the Virasoro-Bott group.
Entire necessary background for the construction of the theory
of unitary representations of $\Diff S^1$ is found in the study of
Kirillov's homogeneous K\"ahlerian manifold $M=\Diff S^1/\Rot S^1$,
where $\Rot S^1$ denotes the group of rotations of $S^1$. The group
$\Diff S^1$ acts as a group of translations on the manifold $M$ with
$\Rot S^1$ as a stabilizer. The K\"ahlerian geometry of $M$ has been
described by Kirillov and Yuriev in \cite{KY1}. The manifold $M$
admits several representations, in particular, in the space of
smooth probability measures, symplectic realization in the space of
quadratic differentials. Let  $A$ stand for  the class of all
analytic regular univalent functions $f$ in $U$ normalized by
$f(0)=0$, $f'(0)=1$. We will use its analytic representation of $M$
 based on the class $\tilde{A}$ of functions from $A$ which
being extended onto the closure $\overline{U}$ of $U$  are supposed
to be smooth on $S^1$. The class $\tilde{A}$ is dense in $A$ in the
local uniform topology of $U$.

Let $\tilde{\Sigma}$ stand for the class of all univalent regular
maps in the exterior $U^*$ of the unit disk $U$ normalized by
$g(\zeta)=c_1\zeta+c_0+c_1\zeta^{-1}+\dots$ which are smooth on $S^1$.
Then, for each $f\in \tilde{A}$ there is an {\it adjoint map} $g\in
\tilde{\Sigma}$ such that $\overline{\mathbb C}\setminus
f(U)=g(U^*)$. The superposition $g^{-1}\circ f$ restricted to $S^1$
is in $M$. Reciprocally, for each element of $M$ there exist such
$f$ and $g$. Observe that a piece-wise smooth closed Jordan curve is
a quasicircle if and only if it has no cusps. So any function $f$
from $\tilde{A}$ has a quasiconformal extension to $U^*$. By this
realization the manifold $M$ is naturally embedded into the
universal Teichm\"uller space $T$. However, defined as a complex
Banach manifold the Teichm\"uller space $T$ requires additional
Hilbert manifold structure to assure the embedding $M\to T$ to
inherit the K\"ahlerian structure of $M$. This has been done by
Takhtajan and Teo in \cite{T3}--\cite{T5}. The K\"ahlerian structure
on $M$ corresponds to the K\"ahlerian structure on $T$ given by the
analogue of the Weil-Petersson metric.

The Goluzin-Schiffer variational formula lifts the actions from the
Lie algebra $\Vect S^1$ onto $\tilde{A}$. Let $f\in\tilde{A}$ and
let $\nu(e^{i\theta})$ be a $C^{\infty}$ real-valued function in
$\theta\in(0,2\pi]$ from $\Vect S^1$ making an infinitesimal action
as $\theta \mapsto \theta+\tau \nu(e^{i\theta})$. Let us consider a
variation of $f$ given by
\begin{equation}
\delta_{\nu}f(\zeta)=-\frac{f^2(\zeta)}{2\pi
i}\int\limits_{S^1}\left(\frac{wf'(w)}{f(w)}\right)^2\frac{\nu
(w)}{f(w)-f(\zeta)}\frac{dw}{w} .\label{var}
\end{equation}
 Kirillov and Yuriev \cite{KY1}, \cite{KY2} have established
that the variations $\delta_{\nu}f(\zeta)$ are closed with respect
to the commutator and the induced Lie algebra is the same as $\Vect
S^1$. Moreover, Kirillov's result \cite{Kir} states that there is
the exponential map $\Vect S^1\to \Diff S^1$ such that the subgroup
$\Rot S^1$ coincides with the stabilizer of the map $f(\zeta)\equiv
\zeta$ from $\tilde{A}$.

\section{Semigroups of conformal maps}

The basic ideas that we use in this section come from the
development of L\"owner's parametric method that emerges at a
seminal L\"owner's paper \cite{Loewner}. L\"owner was first who
proposed to use Lie semigroups of conformal maps to obtain an
evolution equation for conformal maps. His ideas have been furthered
by many authors among whom we mention Pommerenke \cite{Pom1, Pom2}
as a general reference, and  Goryainov's works \cite{Gor, Gor1}
especially closed to our consideration, one also may see
\cite{Shoikhet, Vas1}.

 We consider the semigroup $\mathcal G$ of conformal
univalent maps from $U$ into itself with composition as the
semigroup operation. This makes $\mathcal G$ a topological semigroup
with respect to the topology of local uniform convergence on $U$. We
impose the natural normalization for such conformal maps:
$\displaystyle\Phi(\zeta)=\beta \zeta+b_2\zeta^2 +\dots$,
$\zeta\in U$, $\beta>0$. The unit of the semigroup is the identity.
Let us construct on $\mathcal G$ a one-parameter semi-flow
$\Phi^{\tau}$, that is, a continuous homomorphism from $\mathbb R\sp
+$ into $\mathcal G$, with the parameter ${\tau}\geq 0$. For any
fixed ${\tau}\geq 0$ the element $\Phi^{\tau}$ is from $\mathcal G$
and is represented by a conformal map $\displaystyle
\Phi(\zeta,{\tau})=\beta(\tau)
\zeta+b_2(\tau)\zeta^2+\dots$ from $U$ onto the domain
$\Phi(U,{\tau})\subset U$. The element $\Phi^{\tau}$ satisfies the
following properties:
\begin{itemize}
\item $\Phi^0= id$;

\item $\Phi^{{\tau}+s}=\Phi(\Phi(\zeta,{\tau}),s)$, for $\tau,
s\geq 0$;

\item $\Phi(\zeta,{\tau})\to \zeta$ locally uniformly in $U^*$ as
$\tau\to 0$.

\end{itemize}
In particular, $\beta(0)=1$. This semi-flow is generated by a vector
field $v(\zeta)$ if for each $\zeta\in U$ the function
$w=\Phi(\zeta,\tau)$, $\tau\geq 0$ is a solution of an autonomous
differential equation $dw/d\tau =v(w)$ with the initial condition
$w|_{\tau=0}=\zeta$. The semi-flow can be extended to a symmetric
interval $(-t,t)$ by putting $\Phi^{-\tau}= \Phi^{-1}(\zeta,\tau)$.
Certainly, the latter function is defined on the set $\Phi(U,\tau)$.
Admitting this restriction for negative $\tau$ we define a
one-parameter family $\Phi^{\tau}$ for $\tau\in (-t,t)$.

For a semi-flow $\Phi^{\tau}$ on $\mathcal G$ there is an
infinitesimal generator at ${\tau}=0$ constructed by the following
procedure. Any element $\Phi^{\tau}$ is represented by a conformal
map $\Phi(\zeta,{\tau})$ that satisfies the Schwarz Lemma for the
maps $U\to U$, and hence,
$$\R\frac{\Phi(\zeta,{\tau})}{\zeta}\leq
\Big|\frac{\Phi(\zeta,{\tau})}{\zeta}\Big|\leq 1,\quad \zeta\in U,$$
where the equality sign is attained only for $\Phi^0=id\simeq
\Phi(\zeta,0)\equiv \zeta$. Therefore, the following limit exists
(see, e.g., \cite{Gor, Gor1, Shoikhet})
$$\lim\limits_{{\tau}\to
0}\R\frac{\Phi(\zeta,{\tau})-\zeta}{\tau\zeta}=\R\frac{\frac{\partial
\Phi(\zeta,{\tau})}{\partial {\tau}}\Big|_{{\tau}=0}}{\zeta}\leq
0,$$ and the representation $$\frac{\partial
\Phi(\zeta,{\tau})}{\partial {\tau}}\Big|_{{\tau}=0}=-\zeta
p(\zeta)$$ holds, where $\displaystyle p(\zeta)=p_0+p_1\zeta+\dots$
is an analytic function in $U$ with  positive real part, and
\begin{equation}
\frac{\partial\beta(\tau)}{\partial
\tau}\Big|_{\tau=0}=-p_0.\label{aagenerator}
\end{equation}
In \cite{Gor2} it was shown that $\Phi^{\tau}$ is even $C^{\infty}$
with respect to $\tau$. The function $-\zeta p(\zeta)$ is an
infinitesimal generator for $\Phi^{\tau}$ at ${\tau}=0$, and the
following variational formula holds
\begin{equation}
\Phi(\zeta,\tau)=\zeta-\tau\,\zeta p(\zeta)+o(\tau),\quad
\beta(\tau)=1-\tau p_0+o(\tau).\label{aa1}
\end{equation}
The convergence is thought of as local uniform.  We rewrite
(\ref{aa1}) as
\begin{equation}
\Phi(\zeta,\tau)=(1-\tau p_0)\zeta+\tau\,\zeta(
-p(\zeta)+p_0)+o(\tau)=\beta(\tau)\zeta+\tau\,\zeta(
-p(\zeta)+p_0)+o(\tau).\label{aa2}
\end{equation}

Now let us proceed with the semigroup $\tilde{\mathcal G}\subset
\mathcal G$ of elements from $\mathcal G$ represented by univalent
maps $\Phi$ smooth on $S^1$. By the variation of the identity in
$\tilde{A}$ given by the formula (\ref{var}) we get
\[
\frac{\Phi(\zeta,\tau)}{\beta(\tau)}=\zeta-\frac{\zeta^2}{2\pi
i}\int\limits_{S^1}\frac{d(w)}{w(w-\zeta)}dw+o(\tau),\label{aa3}
\]
for $\nu(w)$ from $\Vect S^1$. Comparing with (\ref{aa2})  we come to the conclusion about $p(\zeta)$:
\begin{equation}
p(z)=p_0+\frac{\zeta}{2\pi
i}\int\limits_{S^1}\frac{\nu(w)}{w(w-\zeta)}dw. \label{aa6}
\end{equation}
The constants $p_0$ and the function $\nu(w)$ must be such that $\R
p(z)>0$ for all $z\in U$.

We summarize these observations in  the following theorem.

\begin{theorem}
Let $\Phi^{\tau}$ be a semi-flow in $\tilde{\mathcal G}$. Then it is
generated by the vector field $v(\zeta)=-\zeta p(\zeta)$,
\[
p(\zeta)=p_0+\frac{\zeta}{2\pi
i}\int\limits_{S^1}\frac{\nu(w)}{w(w-\zeta)}dw,
\]
where $\nu(e^{i\theta})\in \Vect S^1$,  and the holomorphic function
$p(\zeta)$ has positive real part in $U$.
\end{theorem}

This theorem implies that at any point $\tau\geq 0$ we have
\[
\frac{\partial\Phi(\zeta,\tau)}{\partial
\tau}=-\Phi(\zeta,\tau)p(\Phi(\zeta,\tau)).
\]

\section{Evolution families and evolution equations}

A subset $\Phi^{t,s}$ of $\mathcal G$, $0\leq s\leq t$ is called an
{\it evolution family} in $\mathcal G$ if
\begin{itemize}
\item $\Phi^{t,t}= id$;

\item $\Phi^{t,s}=\Phi^{t,r}\circ \Phi^{r,s}$, for $0\leq s\leq
r\leq t$;

\item $\Phi^{t,s}\to id$ locally uniformly in $U^*$ as
$t,s\to\tau$.

\end{itemize}
In particular, if $\Phi^\tau$ is a one-parameter semi-flow, then
$\Phi^{t-s}$ is an evolution family. We consider a subordination
chain of mappings $f(\zeta,t)$, $\zeta\in U$, $t\in [0,t_0)$, where
the function $\displaystyle
f(\zeta,t)=\alpha(t)z+a_2(t)\zeta^2+\dots$ is a analytic univalent
map $U\to \mathbb C$ for each fixed $t$ and $f(U,s)\subset f(U,t)$
for $s<t$. Let us assume that this subordination chain exists for
$t$ in an interval $[0,T)$.

Let us pass to the semigroup $\tilde{\mathcal G}$. So $\Phi^{t,s}$
now has a smooth extension to $S^1$. Moreover, $\Phi^{t,s}\to id$
locally uniformly in $\mathbb C$ as $t,s\to \tau$.

We construct the superposition $f^{-1}(f(\zeta,s),t)$ for $t\in
[0,T)$, $s \leq t$. Putting $s=t-\tau$ we denote this mapping by
$\Phi(\zeta,t,\tau)$.

Now we suppose the following conditions for  $f(\zeta,t)$.
\begin{itemize}

\item[(i)] \ \ The maps $f(\zeta,t)$ form a subordination chain in
$U$, $t\in [0,T)$.

\item[(ii)] \ \ \ The map $f(\zeta,t)$ is holomorphic in $U$,
$f(\zeta,t)=\alpha(t)\zeta+a_2(t)\zeta^2+\dots$, where $\alpha(t)>0$
and differentiable with respect to $t$.

\item[(iii)] \ \ \  The map $f(\zeta,t)$ admits a smooth continuation onto $S^1$.

\end{itemize}

 The function
$\Phi(\zeta,t,\tau)$ is embedded into
 an evolution family in
$\mathcal G$. It is differentiable with regard to $\tau$ and $t$ in
$[0,T)$, and $\Phi(\zeta,t,0)=\zeta$. Moreover, $\zeta=\lim_{\tau\to
0}\Phi(\zeta,t,\tau)$ locally uniformly in $U$ and
$\Phi(\zeta,t,\tau)$ is embedded now into an evolution family in
$\mathcal G^{qc}$. The identity map is embedded into a semi-flow
$\Phi^{\tau}\subset \tilde{\mathcal G}$ (which is smooth) as the
initial point with the same velocity vector
 $$ \frac{\partial
\Phi(\zeta,t,{\tau})}{\partial {\tau}}\Big|_{{\tau}=0}=-\zeta
p(\zeta,t),\quad \zeta\in U, $$ that leads to the L\"owner-Kufarev
equation
\begin{equation}
\dot{f}=\zeta f' p(\zeta,t),\label{LKP}
\end{equation}
 (the semi-flow $\Phi^{\tau}$ is tangent to the evolution
family at the origin). Actually, the differentiable trajectory
$f(\zeta,t)$ generates a pencil of tangent smooth semi-flows with
starting tangent vectors $-\zeta p(\zeta,t)$ (that may be only
measurable with respect to $t$).

Therefore, the conclusion is that the function $f(\zeta,t)$
satisfies the equation (\ref{LKP}) where the function $p(\zeta,t)$
is given by
\[
p(\zeta,t)=p_0(t)+\frac{\zeta}{2\pi
i}\int\limits_{S^1}\frac{\nu(w,t)}{w(w-\zeta)}dw,
\]
and has positive real part. The existence of $p_0(t)$ comes from the
existence of the subordination chain. One may assign the
normalization to $f(\zeta,t)$ controlling the change of the
conformal radius of the subordination chain by, e.g.,  $e^{t}$. In
this case, changing variables we obtain $p_0=1$.

Summarizing the conclusions about the function $p(\zeta,t)$ we come
to the following result.

\begin{theorem} Let $f(\zeta,t)$ be a subordination  chain of
maps in $U$ that exists for $t\in [0,T)$  and satisfies the
conditions (i--iii). Then, there are a real valued function
$p_0(t)>0$ and a real valued function $\nu(\zeta,t)\in \Vect S^1$,
such that $\R p(\zeta,t)>0$ for $\zeta\in U$,
$$p(\zeta,t)=p_0(t)+\frac{\zeta}{2\pi
i}\int\limits_{S^1}\frac{\nu(w,t)}{w(w-\zeta)}dw, \quad \zeta\in U,
$$ and $f(\zeta,t)$ satisfies the L\"owner-Kufarev differential
equation (\ref{LKP}) in $t\in [0, T)$.
\end{theorem}

Comparing (\ref{PG2}) and (\ref{carat}) with this theorem we come to
the following corollary.

\begin{corollary} Let $f(\zeta,t)=\alpha(t)\zeta+a_2(t)\zeta^2+\dots$ be a subordination  chain of
maps in $U$ that parameterizes the classical Laplacian Growth that
exists for $t\in [0,T)$. Then, under the notations of the previous
theorem we have
\[
p_0=\frac{\dot{\alpha}}{\alpha}=\frac{1}{2\pi}\int\limits_{0}^{2\pi}\frac{1}{|f'(e^{i\theta},t)|^2}d\theta,\quad
\mbox{and}\quad \nu(e^{i\theta},t)=\frac{2}{|f'(e^{i\theta},t)|^2}.
\]
\end{corollary}

In the view of this corollary we come to an interpretation of the
variation of the logarithmic action as
\[
\frac{d}{dt}(\mathcal{S}[f]+\mathcal{E}[f])=2\int\limits_{0}^{2\pi}\varkappa^2(\theta,t)\,d\theta+
\R\int\limits_{0}^{2\pi}e^{2i\theta}\nu(e^{i\theta},t) S_f \,\,
d\theta,
\]
where $\nu$ is a vector from  the Lie algebra $\Vect S^1$ tangential
to $\Diff S^1/\Rot S^1$ at the unity.

\section{Connections with the Virasoro algebra}

In two dimensional conformal field theories \cite{Goddard}, the algebra of energy momentum tensor is deformed by a central extension due to the conformal anomaly and becomes the {\it Virasoro algebra}.
The Virasoro algebra is spanned
by elements $e_k=\zeta^{1+k}\partial$, $k\in \mathbb Z$ and $c$ with $e_{k}+e_{-k}$, where $c$ is a real number, called the central charge, and the Lie brackets are defined by
\[
[e_m,e_n]_{Vir}=(n-m)e_{m+n}+\frac{c}{12}m(m^2-1)\delta_{n,-m},\quad [c, L_k]=0.
\]
The Virasoro algebra ($Vir$) can be realized as a central extension
of $\Vect S^1$ by defining 
\[
[\phi\partial+ca , \psi\partial+cb]_{Vir}= (\phi\psi'-\phi'\psi)\partial+\frac{c}{12}\omega(\phi,\psi),
\]
(whereas $[\phi, \psi]= \phi\psi'-\phi'\psi$), where the bilinear antisymmetric form $\omega(\phi,\psi)$ on $\Vect S^1$ is given by
\[
\omega(\phi,\psi)=-\frac{1}{4\pi}\int\limits_{0}^{2\pi}(\phi'+\phi''')\psi d\theta,
\]
and $a,b$ are numbers. This form defines the  Gelfand-Fuks cocycle on $\Vect S^1$ and satisfies the Jacobi identity. The factor of 1/12 is merely a matter of convention.
The manifold $M$ being considered as a realization $\tilde{A}$ admits affine coordinates
$\{c_2,c_3,\dots\}$, where $c_k$ is the $k$-th coefficient of a univalent functions $f\in \tilde{A}$.
Due to de Branges' theorem \cite{Branges}, $M$ is a bounded open subset of $\{|c_k|<k+\varepsilon\}$.

Taking $\nu=-ie^{ik\theta}$, $k\geq 0$, we obtain the expressions for $L_k=\delta_{\nu} f$, 
$f\in \tilde{A}\simeq M$ (see formula (\ref{var})),  as
\[
L_0=\zeta f'(\zeta)-f(\zeta), \quad L_k=\zeta^{1+k}f'.
\]
The computation of $L_k$ for $k<0$ is more difficult because poles of the integrant.
For example,
\[
L_{-1}=f'-1-2c_2f,\quad L_{-2}=\frac{f'}{\zeta}-\frac{1}{f}-3c_2+(c_2^2-4c_3)f,
\]
(see, e.g., \cite{Kir98}). In terms of the coordinates $\{c_2,c_3,\dots\}$ on $M$ 
\[
L_k=\partial_k+\sum\limits_{n=1}^{\infty}(n+1)c_n\partial_{k+n},\quad L_0=\sum\limits_{n=1}^{\infty}nc_n\partial_{n},
\]
for $k>0$, where $\partial_{k}=\partial/\partial c_{k+1}$.

Neretin \cite{Neretin} introduced the sequence of polynomials $P_k$, in the coordinates $\{c_2,c_3,\dots\}$ on $M$ by the following recurrent relations
\[
L_m(P_n)=(n+m)P_{n-m}+\frac{c}{12}m(m^2-1)\delta_{n,m},\quad P_0=P_1\equiv 0,\quad P_k(0)=0,
\]
where the central charge $c$ is fixed. This gives, for example, $P_2=\frac{c}{2}(c_3-c_2^2)$,
$P_3=2c(c_4-2c_2c_3+c_2^3)$. In general, the polynomials $P_k$ are homogeneous with respect to
rotations of the function $f$. It is worthy to mention that estimates of the absolute value of these polynomials
has been a subject of investigations in the theory of univalent functions for a long time, e.g., for $|P_2|$ we have
$|c_3-c_2^2|\leq 1$ (Bieberbach 1916 \cite{Bieberbach}), for estimates of $|P_3|$ see \cite{Gromova, Lehto, Tammi1, Tammi2}. For the Neretin polynomials one can construct the generatrix function  
\[
P(\zeta)=\sum\limits_{k=1}^{\infty}P_k\zeta^k=\frac{c\zeta^2}{12}S_f(\zeta),
\]
where $S_f(\zeta)$ is the Schwarzian derivative of $f$,
Let $\nu\in \Vect S^1$ and $\nu^g$ be the associated right-invariant tangent vector field defined at $g\in \Diff S^1$. For the basis $\nu_k=-ie^{ik\theta}\partial$, one constructs the corresponding  associated right-invariant basis $\nu_k^g$. By $\{\psi_{-k}\}$ we denote the dual basis of 1-forms such that the value
of each form on the vector $\nu_k^g$ is given as
\[
(\psi_k, \nu_n^g)=\delta_{k+n,0}.
\]
Let us construct the 1-form $\Omega$ on $\Diff S^1$ by
\[
\Omega=\sum\limits_{k=1}^{\infty}(P_k\circ \pi)\psi_k,
\]
where $\pi$ means the natural projection $\Diff S^1\to M$.
This form appeared in \cite{Airault, Airault2}  in the context of the construction of a unitarizing probability measure for the Neretin representation of $M$. It is invariant under the left action of $S^1$. If $f\in \tilde{A}$ represents $g$ and $\nu\in \Vect S^1$, then the value of the form $\Omega$ on the vector $\nu$ is
\[
(\Omega, \nu)_{f}=\int\limits_{0}^{2\pi}e^{2i\theta}\nu(e^{i\theta}) S_f \,\,
d\theta,
\]
see \cite{Airault, Airault2}. So the variation of the logarithmic action given in Theorem 1 becomes
\begin{equation}
\frac{d}{dt}(\mathcal{S}[f]+\mathcal{E}[f])=2\int\limits_{0}^{2\pi}\varkappa^2(\theta,t)\,d\theta+
\R (\Omega, \nu)_{f}.\label{Vir}
\end{equation}
In this formula, we take into account  the first coefficient,  the conformal radius of the Laplacian evolution, that does not change the form $\Omega$.

\section{Some open questions}

\begin{itemize}
\item[(i) ] One may conjecture that the formula (\ref{Vir}) remains true for general smooth subordination evolution. 

\item[(ii)] The logarithmic action is clearly related to the universal Liouville action suggested by Takhtajan and Teo in \cite{T3, T4, T5}. There must be possible to obtain the variation given in Theorem 1 by means of the variation of the universal Liouville action obtained in \cite{T4}. This would be yet more interesting  because the universal Liouville action is defined for contours without any smoothness hypothesis. 

\item[(iii)] Another interesting question is whether it is possible to make
regularization of the proper Liouville action integral based on the Poincar\'e metric by the boundary distortion by a univalent function. 

\item[(iv)] A deeper task is concerned with the Laplacian growth and its embedding into
the Whitham-Toda hierarchy. The extended Toda hierarchy (see \cite{Carlet}) admits a nonabelian algebra
of infinitesimal  symmetries isomorphic to half of the Virasoro algebra \cite{Dubrovin}. It would be interesting to reveal the connections between the Hamiltonian approach through the Toda hierarchies
and the action approach suggested in the present paper. 

\item[(v)] The multiply connected Laplacian growth is a natural way of generalization of all these results.
\end{itemize}

\end{document}